\documentclass[journal,onecolumn]{IEEEtran}
\usepackage[utf8]{inputenc}
\usepackage[T1]{fontenc}
\usepackage[english]{babel}
\usepackage{amsmath}
\usepackage{amssymb}
\usepackage[cmintegrals]{newtxmath}
\usepackage{graphicx}
\usepackage{caption}
\usepackage{subcaption}
\usepackage{xcolor}
\usepackage{multirow}
\usepackage{float}
\usepackage{rotating}
\usepackage{mathtools, cuted}

\usepackage{ifpdf}

\usepackage{cite}

\ifCLASSINFOpdf
\else
\fi
\begin{document}
%
\title{Maximal Transmission Rate in Omni-DRIS-Assisted Indoor Visible Light Communication Systems}
%
%
%

\author{Alain R. Ndjiongue, \textit{Senior Member, IEEE}, Octavia A. Dobre, \textit{Fellow, IEEE}, and Hyundong Shin, \textit{Fellow, IEEE} 
\thanks{Copyright (c) 2015 IEEE. Personal use of this material is permitted. However, permission to use this material for any other purposes must be obtained from the IEEE by sending a request to pubs-permissions@ieee.org.}
\thanks{Alain R. Ndjiongue is with the Faculty of Engineering and Applied Science, Memorial University, Canada, and the Faculty of Engineering and the Built Environment, University of Johannesburg, South Africa.} 
\thanks{Octavia A. Dobre (corresponding author) is with the Faculty of Engineering and Applied Science, Memorial University, Canada, and the Department of Electronic Engineering, Kyung Hee University, Yongin 17104, South Korea.}
\thanks{Hyundong Shin (corresponding author) is with the Department of Electronics and Information Convergence Engineering, Kyung Hee University, South Korea.}
}

\maketitle

\begin{abstract}
Given the importance of reconfigurable intelligent surfaces (RISs) in next-generation mobile systems, several RIS variants have been proposed in recent years. Omni-digital-RIS (omni-DRIS) is one of the newly introduced variants of optical RIS that can successfully be driven by bit sequences to control lights emerging from simultaneous reflection and refraction processes, impacting both the achievable rate and the required number of omni-DRIS elements. In this paper, we analyze the effects of omni-DRIS-assisted transmission environment parameters to maximize the achievable rate and highlight the corresponding number of omni-DRIS elements. Furthermore, we show that the number of omni-DRIS elements that yields the highest achievable rate largely depends on the number of bits per omni-DRIS control sequence. On the other hand, this rate is determined by the remaining parameters of the transmission system and environmental factors, which include the total transmit power, transmission bandwidth, number of transmitters and users, and the channel DC gain.
\end{abstract}


%
\IEEEpeerreviewmaketitle

\section{Introduction} \label{intro}
Reconfigurable intelligent surfaces (RISs) offer a number of benefits, including addressing the skip zone problem and providing a degree of control over the transmission environment, to name just a few \cite{9475155, 9475154, 9443170, 9614037, 10138329}. These advantages explain the widespread interest in RIS in research and industry. RIS has been considered to support both radio frequency (RF) and optical wireless communication (OWC) technologies \cite{9475155, 9475154, 9614037, 10138329, 9443170}. The difference between an RIS for RF and an RIS for OWC is the material used in the RIS element, as the material does not respond to all electromagnetic frequencies in the same manner and to the same degree. Several RIS versions have been proposed for OWC systems. These include optical RISs, mirrors, digital RISs (DRISs), and omni-DRISs, to name just four. A DRIS is an optical digital RIS that can be controlled numerically, while an omni-DRIS is a DRIS that can simultaneously reflect and refract the incident light \cite{10138329}. The authors of \cite{10138329} introduced the omni-DRIS concept and described its principles and control code sequences. However, this has not been fully analyzed; e.g., the number of omni-DRIS elements required to achieve a maximal transmission rate has not been discussed. In this paper, we close this gap by analyzing the number of elements that maximize the achievable rate in an indoor visible light communication (VLC) system supported by omni-DRIS.

To connect users located on both sides of the RIS module, simultaneously reflecting and refracting-RIS and omni-DRIS can be used \cite{9437234, 9863732, 10163896, 9849460, 9570143, 10138329}. These RIS variants reflect and/or refract the impinging signal. In addition, they control the amplitude level and the direction of the emerging signals on either side of the RIS module. Further, RIS modules can be digitized. This was the aim of some recent works, such as \cite{9779993}. On the other hand, since data transmission optimization has always been a top priority for telecommunication technologies, optimization works were proposed. For example, this is the case in OWC systems, where the transmission in free-space optical communication and VLC systems was optimized \cite{9837009, 9910023}. In most cases, the optimization involves the RIS selection, the module position, and the orientation of its elements, which can be determined by the yaw and roll angles of the elements \cite{9543660}. In some cases, the optimization is related to the substance or properties of the RIS elements' materials. As an example, in \cite{9910023}, the achievable rate of a VLC system was improved through the refractive index of liquid crystal RIS elements. Furthermore, optimization of RIS-assisted systems based on machine learning techniques was also proposed. For example, \cite{9762646} and \cite{9558821} focused on deep reinforcement learning to optimize full-duplex systems and combined half- and full-duplex systems aided by the RIS technology. Both orientation and location optimization of an RIS-assisted system are provided in \cite{Zeng}, while in \cite{Lim}, optimization of beamforming, phase shift, and association for RIS communication with the base station and the user are provided. The authors of \cite{Faisal} optimized the discrete phase-shift of an RIS-aided system, while in \cite{Lietal}, the system achievable rate considers the transmit power, number of transmission sub-carriers, and number of users. Although all these works aimed at optimizing RIS-assisted systems, none addressed indoor omni-DRIS-assisted VLC systems, where the optimization considers the number of omni-DRIS elements. This is the focus of this paper.

In omni-DRIS, a number, $\zeta_o$, of elements are active and can reflect and refract impinging lights. $\zeta_o$ is less than or equal to the number of omni-DRIS elements. As a result of digitizing the omni-DRIS module, the phase profile and transition coefficient are both digitized. They can be controlled digitally by a series of bits. In practice, the number of omni-DRIS elements is a power of two, $N = 2^k$, where $k$ is the number of bits per control sequence. For instance, the omni-DRIS will have 1, 2, 4, 8, 16, 32, 64, 128, 256, \dots elements. This idea implies that the total number of active omni-DRIS elements may also or may not be a power of two. Thus, it is necessary to find the number of omni-DRIS elements that provide the maximal transmission rate to the system. To the best of our knowledge, this aspect has not been considered in the literature, not even in \cite{10138329}, where the omni-DRIS concept was introduced. Not even in \cite{Yang} where, in addition to the transmit signal-to-noise ratio, the authors also considered the number of RIS elements. Their results show that infinitely increasing the number of RIS elements infinitely increases the sum rate. However, this is not the case in practice because the RIS's size is limited. This highlights a gap in the research literature on RIS-aided optical systems. We think of increasing the number of elements of an omni-DRIS module while preserving its size. Therefore, the purpose of this paper is to fill the gap by studying the number of omni-DRIS elements that yields an enhanced transmission rate in an omni-DRIS-assisted indoor VLC system.
\begin{figure*}
	\centering
	\includegraphics[width=1\textwidth]{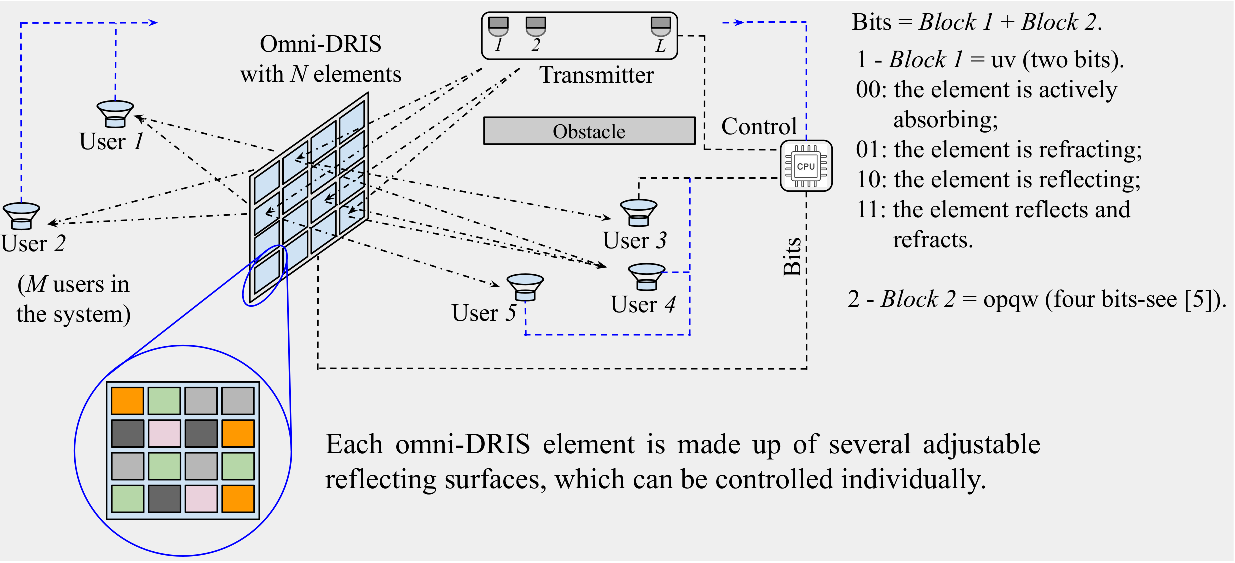}
	\caption{A system model of omni-DRIS-assisted indoor VLC system.}
	\label{fig:model}
\end{figure*}
The main contributions of this paper are enumerated as follows:
\begin{enumerate}
    \item We discuss the achievable rate of an omni-DRIS assisted indoor VLC system in which users are positioned on both sides of the RIS module.
   \item We analyze the system considering the number of omni-DRIS elements by exploiting the Newton-Mercator power series expansion. 
   \item We demonstrate that the maximum transmission rate upper-bound relates to the number of omni-DRIS elements.
\end{enumerate}
To obtain the maximal achievable rate, we discuss the effects of the system parameters on the transmission rate. We model the achievable rate as a function of the variable $N$, which represents the number of omni-DRIS elements. The remaining system parameters like bandwidth, number of transmitters and users, and noise power determine the achievable rate. Subsequently, we derive the number of bits per control sequence as a function of these parameters and the achievable rate.
\section{Omni-DRIS-Assisted Indoor VLC System Model}
Consider the indoor VLC system illustrated in Fig.~\ref{fig:model}. It consists of a transmitter made of $L$ light sources (LSs), an omni-DRIS module consisting of $N$ elements, and $M$ users. The number and orientation of emerging lights are determined by the number of elements in the omni-DRIS module. We assume that the omni-DRIS size is constant. Therefore, increasing the number of omni-DRIS elements implies that the size of these elements reduces. This scenario corresponds to the situation where each omni-DRIS element is made up of several adjustable reflecting or refracting surfaces, which can be controlled individually, thus being considered as elements. LSs are made of a light-emitting diode each and are efficiently organised to provide maximal lighting for both the environment and the omni-DRIS module. With the assumption that light reflected from the wall is neglected, the LSs generate the necessary power to carry the message to the $M$ users via non-line of sight (NLoS) paths through the omni-DRIS module. The elements of the omni-DRIS module can reflect and refract light simultaneously, depending on the control unit requirements dictated by the users' location. For example, if all users are on the front side of the omni-DRIS module, all elements should be configured as reflectors. If they are all located behind the module, all elements should be refractors. When users are located on both sides of the omni-DRIS module, the control unit configures the elements based on the amount of energy required to power these users. To achieve a trade-off power between the two sides of the omni-DRIS module, some elements may be set to act simultaneously as reflector and refractor. Note that all these elements have been optimized with transition coefficients close to unity. This coefficient is shared between reflection and refraction in the event that the element is required to reflect and refract the incident light. The users' receiving system utilizes liquid crystals and/or lenses to improve their field-of-view and eliminate most unwanted and background lights. Two main types of noise are considered at users: thermal and shot noises. These two types of noise are assumed to be additive white Gaussian noises, with zero means and variances given by $\Lambda^r_0/2$, with $\Lambda^r_0$ denoting the power spectral density (PSD) of the resultant noise at the user \cite{7208804}. Therefore, the noise at the user is defined as $n \sim \mathcal{N} \big(0, \Lambda^r_0/2 \big)$.

An illustration of the mapping technique and mapping table for a coding scheme for the omni-DRIS analyzed in this paper is provided at the bottom part of Fig.~2 of \cite{10138329}. Since this development is not part of our analysis, we avoid repeating it in our paper. The omni-DRIS contains up to 16 different element types. The first part of the mapping is made of two bits defining the element mode that induces the physical process (absorption, reflection, and/or refraction). The other part of the code characterizes the phase profile and transition coefficient (see Fig.~\ref{fig:model}) \cite{10138329}. To facilitate the omni-DRIS control, the elements are grouped per type, and those of the same type receive the same control code.\footnote{Elements of the same type will reflect and/or refract light with the same phase shift and transition coefficient.}
\section{Analysis and Results}
\subsection{Analysis}
Consider the $L$ LSs of the transmitter and the $N$ elements of the omni-DRIS module, with $\vartheta$ elements actively absorbing, $\zeta_o = N - \vartheta \leq N = 2^k $. Also consider the $M$ users equipped with a single photodetector each. Assuming that the LoS links are completely obstructed for users on the front side of the omni-DRIS module (see Fig.~\ref{fig:model}), the achievable rate of a message originated from the $l$-th LS, traveling through the $n$-th omni-DRIS element, and landing on the $m$-th user's photodetector can be expressed as \cite{9714890, 9615188}
\begin{equation}
    R_{o} = \frac{1}{2} W \log_2\left(1+\frac{e}{2\pi} \gamma_{o}\right),
    \label{Eq:rate0}
\end{equation}
where $W$ denotes the signal bandwidth, $e = 2.718$ is the base of the Napierian logarithm, and $\gamma_{o}$ is the signal-to-noise ratio of the NLoS link $l$-th LS-$n$-th omni-DRIS element-$m$-th-user, expressed as
\begin{equation}
    \gamma_{o} = \frac{\rho^2\left[\frac{P_t}{MNL}G_{NLoS}\right]^2}{\Lambda_0/2},
    \label{Eq:SNR}
\end{equation}
with $\rho$ as the optical-to-electrical conversion coefficient at the user. $P_t$ represents the total transmit power distributed to the $L$ LSs, $N$ omni-DRIS elements, and $M$ users, and the quantity $P_t/(MNL)$ denotes the portion of the total transmit power attributed to a single LS illuminating a single user through an active omni-DRIS element. $\Lambda_0$ represents the noise PSD at the user considering the link $l$-th LS-$n$-th omni-DRIS element-$m$-th user, and $G_{NLoS}$ is the corresponding DC channel gain, expressed as \cite{9543660}
\begin{equation}
\begin{split}
            G_{NLoS} = & \frac{\eta_n A_oA_{pd}(r + 1)}{2\pi (d_{l,n})^2(d_{n,m})^2} \cos^r(\theta_{l,n}) \cos(\phi_{l,n}) \cos(\theta_{n,m}) \\ & \times \cos(\phi_{n,m}) T(\phi_{n,m}) g(\phi_{n,m}),
\end{split}
    \label{gain_N}
\end{equation}
where $r$ denotes the Lambertian order, $\eta_n$ is the reflectiveness of the $n$-th active omni-DRIS element, and $A_o$ and $A_{pd}$ are the effective areas of the $n$-th omni-DRIS element and photodetector, respectively. $d_{l,n}$ is the distance between the $l$-th LS and the $n$-th active omni-DRIS element; $d_{n,m}$ is the distance from the $n$-th omni-DRIS active element to the $m$-th user. $\theta_{l,n}$ and $\phi_{l,n}$ are the angles of irradiance from the $l$-th LS toward the $n$-th active omni-DRIS element and from the $n$-th active omni-DRIS element toward the user, respectively. $\theta_{n,m}$ and $\phi_{n,m}$ are the angles of incidence on the $n$-th omni-DRIS reflective/refractive surface and the user, respectively, and $T(\phi_{n,m})$ and $g(\phi_{n,m})$ are the optical concentration and optical filter gains, respectively.

For $L$ LSs, $\zeta_o = N - \vartheta$ active reflective/refractive omni-DRIS elements, and $M$ users, the total achievable rate of the system, $\mathcal{R}$, can be expressed as
\begin{equation}
    \mathcal{R} = \sum_{l=1}^L \sum_{m=1}^M \sum_{n=1}^{N - \vartheta} R_{o}.
    \label{ratess}
\end{equation}

$N$ omni-DRIS elements are assumed to be illuminated, including those actively absorbing. Considering that $R_{o}$ is the same for all active links through reflection and refraction, \eqref{ratess} reduces to 
\begin{equation}
    \mathcal{R} = \frac{WML(N - \vartheta))}{2} \log_2\left(1+\frac{e}{2\pi} \frac{\rho^2 G^2_{NLoS}}{\Lambda_0/2} \left(\frac{P_t}{MNL}\right)^2\right).
    \label{ratess1}
\end{equation}
With $\zeta_o = N - \vartheta$ active reflective/refractive omni-DRIS elements, $(N - \vartheta) \times P_t/MNL$ lights emerge from the omni-DRIS module. Based on \eqref{ratess1}, the number of bits per control code can be derived as
\begin{equation}
    k = \frac{1}{2} \log_2\left(\frac{{\alpha}}{{\psi}\mathrm{e}^\frac{\ln\left(2\right)\,R}{{\xi}{\zeta}_\text{o}}-{\psi}}\right),
\end{equation}
where $\xi = WLM/2$, $\psi = (ML)^2$, and 
\begin{equation*}
    \alpha = \frac{e}{2\pi} \frac{\rho^2 G^2_{NLoS}P_t^2}{\Lambda_0/2}.
    \label{Eq:a}
\end{equation*}
\noindent
It can be seen that \eqref{ratess1} is of the form 
\begin{equation}
    f(N) = \xi(N - \vartheta)\log_2\left(\frac{\alpha}{N^2\psi}+1\right).
    \label{Eq:f(x)}
\end{equation}
\noindent
In practice, $N \gg 1$. Thus, $0 < \frac{\alpha}{N^2\psi} \leq 1$, $\forall$ $N \in \mathbb{N}_{+} - [0, 1[$ holds. Therefore, the Newton-Mercator power series expansion of $\ln[\alpha/(N^2\psi) + 1]$, i.e. \cite[p. 33]{havil2003} 
\begin{equation*}
  \begin{split}
    \ln \left(\frac{\alpha}{N^2\psi}+1\right) & = \sum_{n = 1}^{\infty} \left(-1\right)^{n+1} \frac{\left(\frac{\alpha}{N^2\psi}\right)^n}{n} \\
   & = \left[\frac{\alpha}{N^2\psi}-\frac{\alpha^2}{2N^4\psi^2}+\frac{\alpha^3}{3N^6\psi^3} - \frac{\alpha^4}{4N^8\psi^4} +  \dots\right],
     \end{split}
\end{equation*}
can be exploited to evaluate $f(N)$. Thus, $f(N)$ can be given by
\begin{equation}
    f(N) = \frac{\xi(N-\vartheta)}{\ln(2)}\left[\frac{\alpha}{N^2\psi}-\frac{\alpha^2}{2N^4\psi^2}+\frac{\alpha^3}{3N^6\psi^3} - \dots\right].
    \label{Eq:fapprox}
\end{equation}
The maximal values of the achievable rate are obtained for values of $N$ that yield local maxima of the function $f(N)$. To obtain these local inflection values, we consider two terms of the power series and solve 
\begin{equation}
\begin{split}
& \frac{\partial}{\partial N}\left[f(N)\right] = 0 \\
& \Longrightarrow \frac{{\xi} \left(\frac{2{\alpha}^2}{{\psi}^2N^5}-\frac{2{\alpha}}{{\psi}N^3}\right)\left(N-\vartheta \right)}{\ln\left(2\right)}+\frac{{\xi} \left(\frac{{\alpha}}{{\psi}N^2}-\frac{{\alpha}^2}{2{\psi}^2N^4}\right)}{\ln\left(2\right)} = 0.
\end{split}
    \label{Eq:derive} 
\end{equation}
Equation \eqref{Eq:derive} can be further reduced to 
\begin{equation}
\frac{{\alpha}{\xi}\left(2{\psi}N^3-4{\psi}\vartheta N^2-3{\alpha}N+4{\alpha}\vartheta \right)}{2\ln\left(2\right)\,{\psi}^2N^5} = 0.
\label{Eq:XX} 
\end{equation}
The expression $2{\psi}N^3-4{\psi}\vartheta N^2-3{\alpha}N+4{\alpha}\vartheta$ has three roots, which depend on the values of the parameters $\alpha$, $\psi$, $\xi$, and $\vartheta$. The meaningful root can be approximated as
\begin{equation}
    \begin{split}
    N \approx 
        &\left[\frac{\sqrt{-\frac{{\alpha}\cdot\left(128{\psi}^2\vartheta^4+18{\alpha}{\psi}\vartheta^2+27{\alpha}^2\right)}{{\psi}}}}{6^\frac{3}{2}{\psi}}+\frac{8\vartheta^3}{27}+\frac{\frac{6{\alpha}\vartheta}{2{\psi}}-\frac{6\vartheta{\alpha}}{{\psi}}}{6}\right]^{\frac{1}{3}}\\
        &-\frac{-\frac{4\vartheta^2}{9}-\frac{3{\alpha}}{6{\psi}}}{\left[\frac{\sqrt{-\frac{{\alpha} \left(128{\psi}^2\vartheta^4+18{\alpha}{\psi}\vartheta^2+27{\alpha}^2\right)}{{\psi}}}}{6^\frac{3}{2}{\psi}}+\frac{8\vartheta^3}{27}+\frac{\frac{6{\alpha}\vartheta}{2{\psi}}-\frac{6\vartheta{\alpha}}{{\psi}}}{6}\right]^{\frac{1}{3}} }+\frac{2\vartheta}{3}.
    \end{split}
\end{equation} 
\subsection{Results}
\begin{figure}[b]
	\centering
	\includegraphics[width=0.5\textwidth]{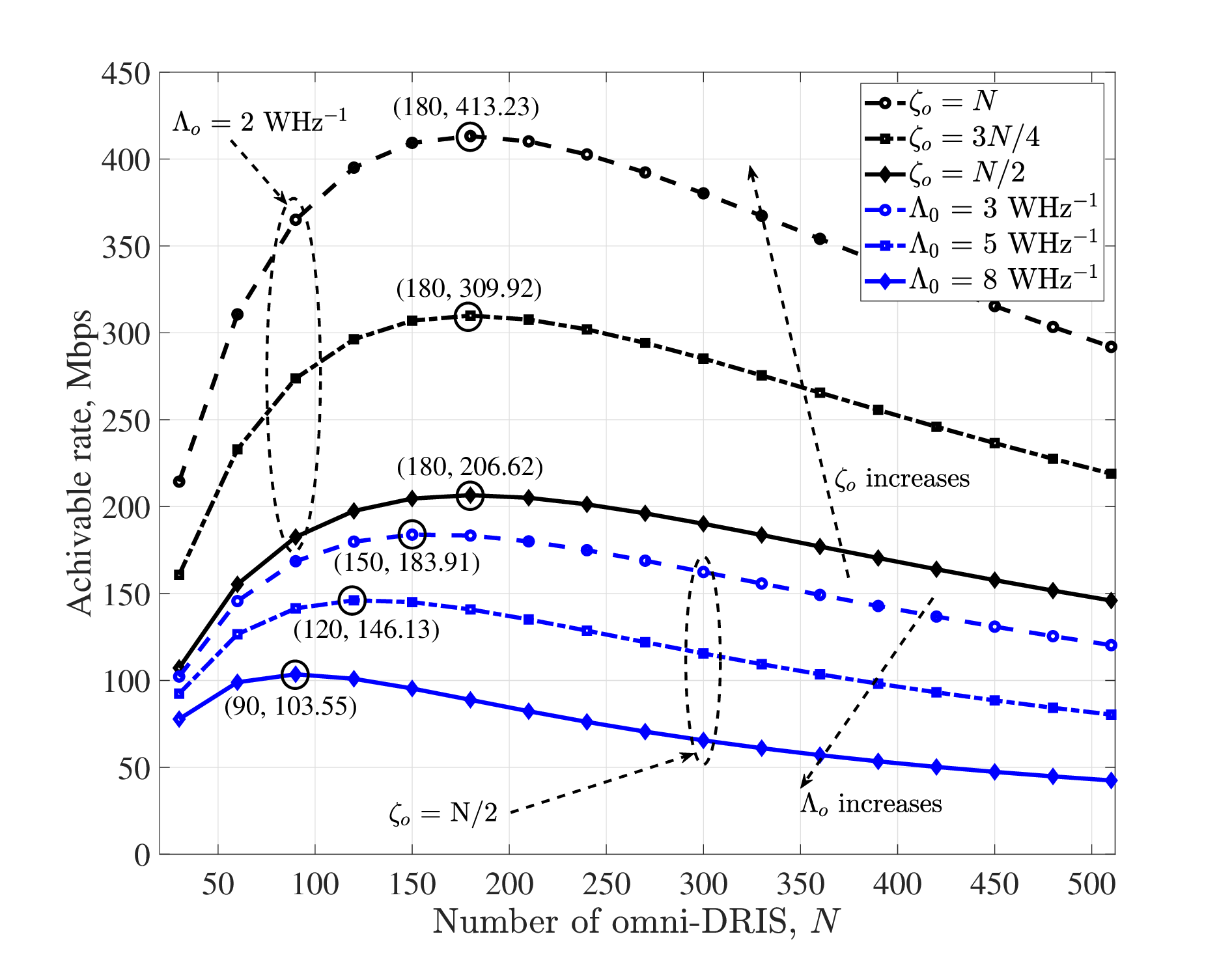}
	\caption{Analytical achievable rate of an omni-DRIS assisted indoor VLC system vs. the number of omni-RIS elements for different numbers of active elements and noise levels, respectively, and a single LS. The values of $N$ for maximal values of the achievable rate are highlighted.}
	\label{fig:2}
\end{figure}
In this paper, we adopt the parameter values and the tetrahedron disposition of users according to the literature \cite{10138329}. The users are located on both sides of the omni-DRIS module and characterised by their azimuth and elevation angles, given as A(31.22$^{\text{o}}$, -27.39$^{\text{o}}$), B(-31.22$^{\text{o}}$, -27.38$^{\text{o}}$), C(31.22$^{\text{o}}$, 27.39$^{\text{o}}$), and D(-31.22$^{\text{o}}$, 27.39$^{\text{o}}$) on one side of the omni-DRIS module, and A'(36.47$^{\text{o}}$, -30.33$^{\text{o}}$), B'(36.47$^{\text{o}}$, 30.33$^{\text{o}}$), C'(-36.47$^{\text{o}}$, 30.33$^{\text{o}}$), and D'(-36.47$^{\text{o}}$, 30.33$^{\text{o}}$) on the other side. These values are derived based on the distances LSs-omni-DRIS, $d_{l,n} = 1.52$ m and omni-DRIS-user, $d_{n,m} = 2.03$ m \cite{10138329}. The bandwidth is set to $W$ = 1 MHz, $1 \leq N \leq 512$ ($N = 2^k$), and the total transmit power is set to $P_t = 10$ W. Up to four lights can be switched on at the transmitter, $1 \leq L \leq 4$, and up to eight users are allowed in the environment, $1 \leq M \leq 8$. The noise is characterized by its common PSD, $\Lambda_0 \in $ \{1, 2, 3, 4, 5, 6, 7, 8, 9, 10\} WHz$^{-1}$. The omni-DRIS effective area is $A_{o}$ = 0.04 m$^2$ and that of each photodetector is $A_{pd} = 4 \times 10^{-4}$ m$^2$. The Lambertian order, $r$, is unity. The angles of irradiance are $\theta_{l,n}$ = 45$^o$, $\phi_{l,n}$ = 10$^o$, and the angles of incidence are $\theta_{n,m}$ = 17.95$^o$, and $\phi_{n,m}$ = 29.58$^o$. Note that the filters are defined as $T(\phi_{n,m})$ = $g(\phi_{n,m})$ = 1, $\eta_n$ = 0.5 for all elements, and $\rho = 0.5$ for all users. 

Figure~\ref{fig:2} depicts the system's achievable rate obtained for $M = 1$ and $L = 1$, using the parameters listed above. The curves are organized in two main groups. The first group is determined by $\Lambda_0 \in $ \{3, 4, 8\} WHz$^{-1}$, when half of the omni-DRIS elements ($\zeta_o = N/2$, $\vartheta = N/2$) are used in the reflection and refraction processes. This is represented in the figure by the three curves at the bottom. The second group of curves is obtained for $\Lambda_0 = 2$ WHz$^{-1}$, while varying the number of active omni-DRIS elements as $\zeta_o = N$, $\zeta_o = 3N/4$, and $\zeta_o = N/2$. This is illustrated by the three curves at the top of Fig.~\ref{fig:2}.

In both groups, it is clear that the highest achievable rate is obtained for a specific number of omni-DRIS elements, depending on the varying parameters and keeping the other parameters constant. Since the number of omni-DRIS elements is a power of two, we select the nearest $2^k$ value that yields the highest achievable rate. As underlined in Table~\ref{tab:2}, we select between the two $2^k$ values before and after the obtained value of $N$, if it is not a power of two. When varying the number of active elements, the highest achievable rate occurs at the same value of $N$, as shown by the three top curves in Fig.~\ref{fig:2}. The markers indicate $N = 180$, for $\zeta_o = N$, $\zeta_o = 3N/4$, and $\zeta_o = N/2$. In this case, the selected value of $N$ is $N$ = $N'$ = 128, leading to $\zeta_o =  128$, $\zeta_o = 96$, and $\zeta_o = 64$ for $\zeta_o = N$, $\zeta_o = 3N/4$, and $\zeta_o = N/2$, and $\vartheta = 0$, 32, and 64, respectively. Accordingly, the maximal achievable rates are 399.59 Mbps, 299.69 Mbps, and 199.90 Mbps, respectively. When varying the noise level, $\Lambda_0 \in $ \{3, 5, 8\} WHz$^{-1}$, the number of omni-DRIS elements that leads to the highest achievable rate also varies, as shown on the three bottom curves of Fig.~\ref{fig:2}. We obtain the maximal achievable rate at $N = 150$, 120, and 90 elements for $\Lambda_0 = 3$ WHz$^{-1}$, 4 WHz$^{-1}$, and 8 WHz$^{-1}$. In this case, the selected values of $N$ are 128 ($N'$), 128 ($N''$), and 64 ($N'$), and the corresponding achievable rates are 181.34 Mbps, 146.27 Mbps, and 99.94 Mbps, with $\zeta_o = \vartheta$ = 64, 64, and 32, respectively.
\begin{figure}
	\centering
	\includegraphics[width=0.5\textwidth]{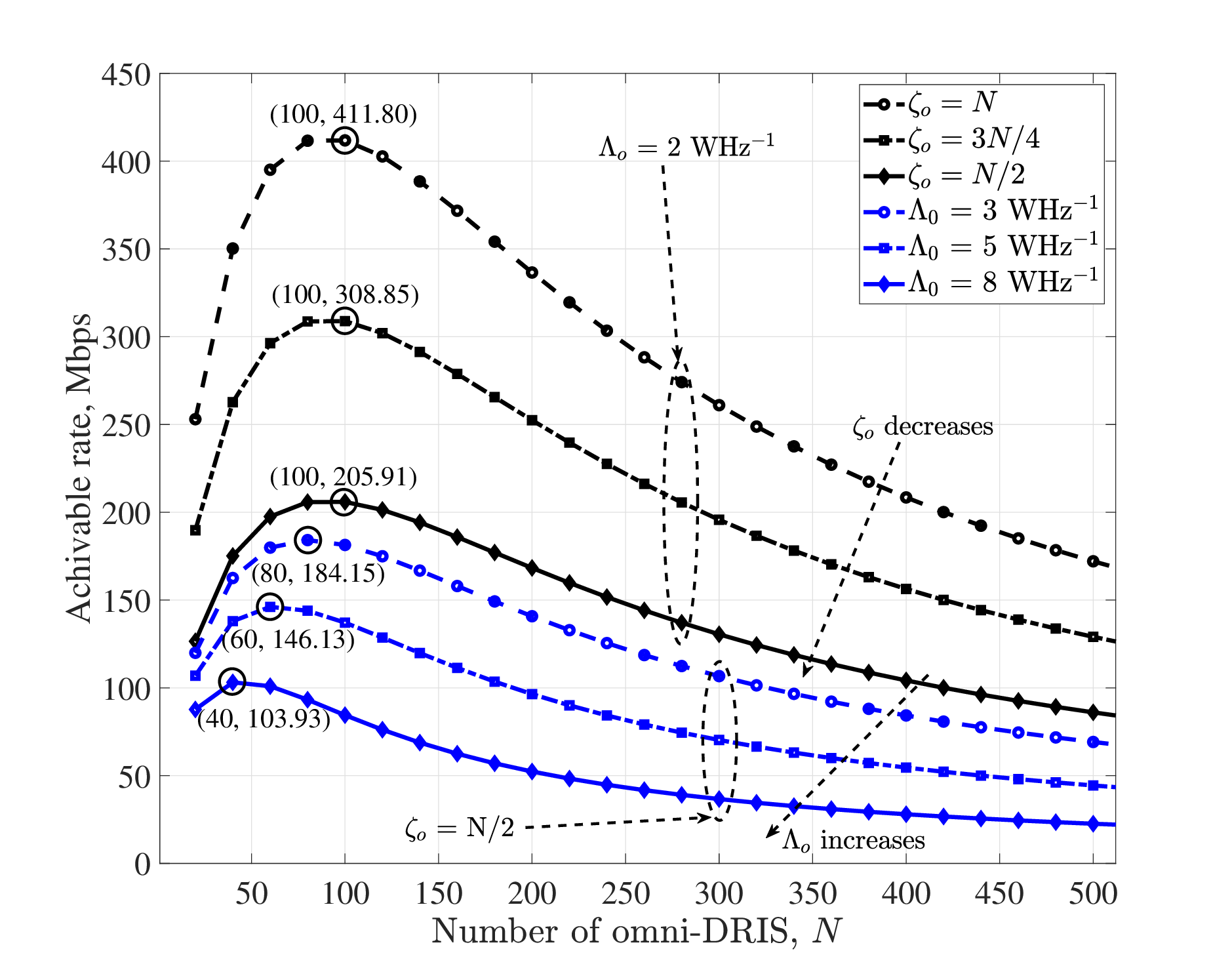}
	\caption{Analytical achievable rate of an omni-DRIS assisted indoor VLC system vs. the number of omni-RIS elements for different numbers of active elements and noise levels, respectively, and two LSs. The values of $N$ for maximal values of the achievable rate are highlighted.}
	\label{fig:4}
\end{figure}
\begin{table}
    \centering
       \caption{Selected vs. Obtained Values of $N$.}
    \label{tab:2}
    \begin{tabular}{c|l|c|c|c}
    \hline
  &  & $N'$-Rate & $N$-Rate & $N''$-Rate \\ \hline \hline
 \multirow[c]{3}{*}[0in]{$\Lambda_0$}
  & $\zeta_o = N$ & \underline{128-399.59} & 180-413.23 & 256-397.76 \\
  & $\zeta_o = 3N/4$ & \underline{128-299.69} & 180-309.92 & 256-298.32 \\
  & $\zeta_o = N/2$ & \underline{128-199.80} & 180-206.62 & 256-198.88 \\ \hline
\multirow{3}{*}{$\zeta_o$} 
  & $\Lambda_0 = 3$ & \underline{128-181.34} & 150-183.91 & 256-172.01 \\
  & $\Lambda_0 = 5$ & 64-128.75 & 120-146.13 & \underline{128-146.27} \\
  & $\Lambda_0 = 8$ & \underline{64-99.94} & 90-103.55 & 128-99.81 \\ \hline
  \multicolumn{5}{c}{$\Lambda_0$ = 2 $WHz^{-1}$, $\zeta_o = N/2$, and}\\
  \multicolumn{5}{c}{$N'$ and $N''$ are $2^k$ values immediately before and after $N$.} \\
 \hline
    \end{tabular}
\end{table}

Figure~\ref{fig:4} depicts the system's achievable rate obtained for $M$ = 1 and $L$ = 2, using the same parameters as in Fig.~\ref{fig:2}. Compared to Fig.~\ref{fig:2}, Fig.~\ref{fig:4} shows the same pattern, with the difference that increasing the number of LSs reduces the number of omni-DRIS elements required to achieve a similar maximal rate. For example, with a single LS, we obtain 399.59 Mbps with 128 elements, while with two LSs, we reach 398.99 Mbps with 64 elements when $\zeta_o = N$. Also note that all the inflection points in Figs.~\ref{fig:2} and \ref{fig:4} are corroborated by the solutions of \eqref{Eq:XX}.
\subsection{Effects of $\alpha$, $\psi$, $\vartheta$, and $\xi$ on $N$ and $f(N)$}
Some of the system parameters that may influence the achievable rate and the corresponding number of omni-DRIS elements, have not been investigated in the above analysis. Four key parameters appear in \eqref{Eq:f(x)}, namely $\alpha$, $\vartheta$, $\xi$, and $\psi$. They reflect the dependency on the bandwidth, number of LSs, number of omni-DRIS elements, number of users, total transmit power, channel DC gain, and noise power. Using \eqref{Eq:f(x)}, we present an analysis of the impact of $\alpha$, $\psi$, $\vartheta$, and $\xi$ on the maximal achievable rate.

\begin{figure}
	\centering
	\includegraphics[width=0.5\textwidth]{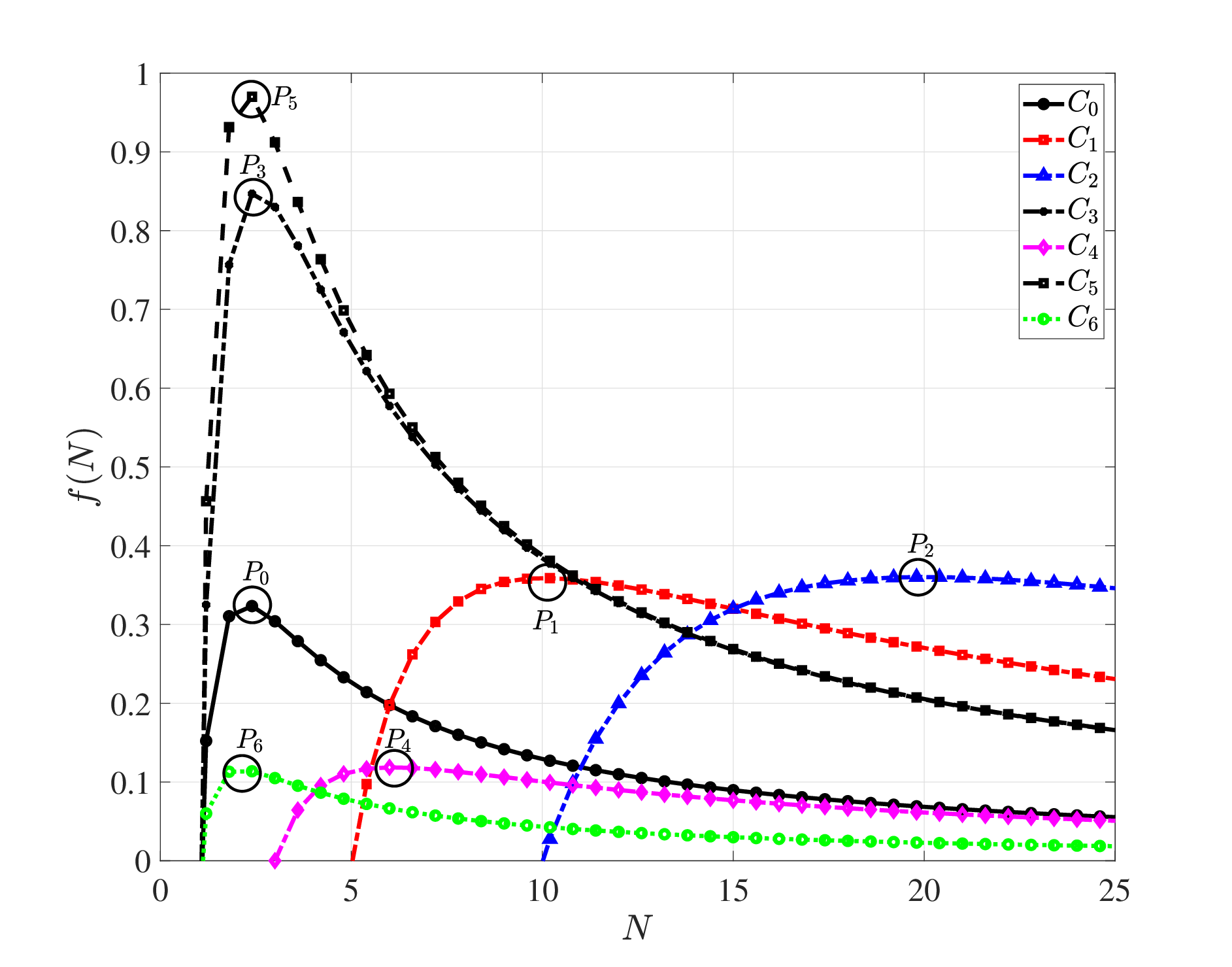}
	\caption{$f(N)$ with normalized values of $\alpha$, $\psi$, $\vartheta$, and $\xi$.}
	\label{fig:3}
\end{figure}

Figure~\ref{fig:3} depicts $f(N)$ in \eqref{Eq:f(x)} for different combinations of the above-mentioned key parameters. These combinations are $C_0$: $\alpha = \vartheta = \xi = \psi$ = 1; $C_1$: $\alpha = \vartheta = \xi = \psi$ = 5; $C_2$: $\alpha = \vartheta = \xi = \psi$ = 10; $C_3$: $\alpha = 3$, $\vartheta = \xi = \psi$ = 1; $C_4$: $\vartheta = 3$, $\alpha = \xi = \psi$ = 1; $C_5$: $\xi = 3$, $\alpha = \vartheta = \psi$ = 1; and $C_6$: $\psi = 3$, $\alpha = \vartheta = \xi$ = 1. The curves demonstrate the impact of all system parameters on the achievable rate, $f(N)$, and the number of omni-DRIS elements, $N$. In all cases, the curves show that at certain values of $N$, the system reaches a maximal achievable rate. However, varying the parameters $\alpha$, $\vartheta$, $\xi$, and $\psi$ does not impact the system performance in the same manner. By setting $C_0$ as our benchmark combination, we make two observations as follows.

When compared to $C_0$, the combinations $C_3$, $C_5$, and $C_6$ do not significantly change the value of $N$ that yields the local maxima of the curves. For $C_0$, the maximum is obtained at $N = 2.2$ ($P_0$), while for $C_3$, $C_5$, and $C_6$, this is observed at $N$ = 2.5 ($P_3$), 2.2 ($P_5$), and 2.1 ($P_6$), respectively. In these three combinations, namely $C_3$, $C_5$, and $C_6$, $\alpha$, $\xi$, and $\psi$ change the values between 1 and 3 with respect to their values in $C_0$, for a limited variation of the value of $N$. Thus, it is legitimate to say that varying $\alpha$, $\xi$, or $\psi$, while keeping the other parameters unchanged, does not change the value of $N$ that provides the local maxima. However, they impact the value of $f(N)$. For example, $C_0$ leads to 0.3252, while $C_3$, $C_5$, and $C_6$ yield 0.8484, 0.9755, and 0.1156. Since $\alpha$, $\xi$, and $\psi$ are determined by parameters such as bandwidth, total transmit power, noise power, channel DC gain, and number of LSs and users, varying these parameters will influence the maximal value of the achievable rate. 

When compared to $C_0$, the combinations $C_1$, $C_2$, and $C_4$ significantly change the value of $N$ that provides the highest achievable rate. While its value is 2.2 ($P_0$) for $C_0$, it is 10 ($P_1$), 20 ($P_2$), and 6.1 ($P_4$) for $C_1$, $C_2$, and $C_4$, respectively. Additionally, in these combinations, $\vartheta$ changes values from one combination to another. $\vartheta$ = 1, 5, 10, 3 for $C_0$, $C_1$, $C_2$, and $C_4$, respectively. Thus, $\vartheta$ is the main parameter that modifies the value of $N$ for a maximal achievable rate, being related to the number of bits per control sequence in the system. Thus, varying this parameter considerably impacts the number of omni-DRIS elements that generate the highest achievable rate. To show the accuracy of our derivation, Table~\ref{tab:table1} depicts values of $N$ and $f(N)$ obtained by simulation against the corresponding calculated values for $C_0$ to $C_6$.  

\begin{table}
  \begin{center}
    \caption{Normalized Values of $N$ and $f(N)$: Simulated vs. Calculated Values.}
    \label{tab:table1}
    \begin{tabular}{c|r|r|r|r}
    \hline
     Scenario & Meas. $N$ &  Calcul. $N$ & Meas. $f(N)$ & Calcul. $f(N)$ \\ \hline \hline
      $C_0$ & 2.2000 & 2.2728 & 0.3252 & 0.3211   \\
      $C_1$ & 10.000 & 10.0502 & 0.3588 & 0.3589   \\
      $C_2$ & 20.0008 & 20.0250 & 0.3602 & 0.3602   \\
      $C_3$ & 2.5000 & 2.8406 & 0.8484 & 0.8037   \\
      $C_4$ & 6.1000 & 6.0845 & 0.1186 & 0.1186   \\
      $C_5$ & 2.2000 & 2.8406 & 0.9755 & 0.9632   \\
      $C_6$ & 2.1000 & 2.0865 & 0.1156 & 0.1154   \\ \hline
      \multicolumn{5}{c}{Meas. = measured and Calcul. = calculated. }\\
 \hline
    \end{tabular}
  \end{center}
\end{table}

\section{Conclusion}
This paper examined the impact of omni-DRIS-assisted indoor VLC system parameters on both the number of omni-DRIS elements and the maximal achievable rate. An omni-DRIS module was integrated into an indoor VLC to resolve the LoS signal blockage and improve data transmission. According to our findings, the maximum transmission rate has an upper bound, which relates to the number of omni-DRIS elements. This number of elements is mainly determined by the number of bits per control sequence rather than other system parameters. However, the remaining parameters of the system determined the maximal achievable rate.

\ifCLASSOPTIONcaptionsoff
\fi
\bibliographystyle{IEEEtran}
 \bibliography{OmniMax}

\end{document}